\documentclass[aps,prc,showpacs,twocolumn,preprintnumbers,floatfix,
showkeys,tightenlines]{revtex4}
\usepackage{amsmath,amssymb,amsfonts}
\usepackage{graphicx}
\usepackage{color}
\allowdisplaybreaks

\newcommand{\be}{\begin{equation}}
\newcommand{\ee}{\end{equation}}
\newcommand{\bea}{\begin{eqnarray}}
\newcommand{\eea}{\end{eqnarray}}
\newcommand{\bm}{\bibitem}

\begin{document}



\title{ Temperature dependence of volume and surface 
 symmetry energy coefficients of nuclei}

\author{J. N. \surname{De}}
\email{jn.de@saha.ac.in}
\author{S. K. \surname{Samaddar}}
\email{santosh.samaddar@saha.ac.in}
\author{B. K. \surname{Agrawal}}
\email{bijay.agrawal@saha.ac.in}
\affiliation{
Saha Institute of Nuclear Physics, 1/AF Bidhannagar, Kolkata
{\sl 700064}, India} 


\begin{abstract} 

The thermal evolution of the energies and free energies of a set
of spherical and near-spherical nuclei spanning the whole periodic
table are calculated in the subtracted finite-temperature Thomas-
Fermi  framework with the zero-range Skyrme-type KDE0 and
the finite-range modified Seyler-Blanchard interaction. The 
calculated energies are subjected to a global fit in the spirit
of the liquid-drop model. The extracted parameters in this model
reflect the temperature dependence of the volume symmetry and surface
symmetry coefficients of finite nuclei, in addition to that of the
volume and surface energy coefficients. The temperature dependence 
of the surface symmetry energy is found to be very substantial
whereas that of the volume symmetry energy turns out to be 
comparatively mild. 
\end{abstract}



\maketitle

The symmetry energy coefficient $a_{sym}^v$ of infinite nuclear matter
is conventionally defined by the relation $e(X)=e(X=0)+a_{sym}^v X^2$.
Here $e$ is the energy per nucleon of the system at isospin asymmetry
$X=(\rho_n-\rho_p)/(\rho_n+\rho_p),~\rho_n $ and $\rho_p$ being the
neutron and proton densities, respectively of the system. For homogeneous
nuclear matter, this definition works extremely well,  $e(X)$ is 
seen to be bilinear in $X$ for nearly all values of asymmetry
\cite{xu,de1}. For warm nuclear matter, the symmetry free energy
coefficient $f_{sym}^v$ is likewise obtained from $f(X,T)-f(X=0,T)=
f_{sym}^v(T) X^2$, where $f(X,T)$ is the per-nucleon free energy
of the matter at asymmetry $X$ and temperature $T$. These asymmetry
coefficients are measures of the energy or free energy release
in converting asymmetric nuclear system to a symmetric one.
For infinite nuclear systems at saturation density $\rho_0$ and
temperature $T=0$, the value of $a_{sym}^v$ is usually taken in
the range of $\sim $ 30-34 MeV \cite{mye,dan,jia}.

In the global fitting of the nuclear masses in the framework of
the liquid-drop mass formula, the symmetry coefficient $a_{sym}$
enters as a phenomenological parameter. Nuclei being finite
systems, it is realized that varying density profiles of different
nuclei necessitate introduction of a mass-dependent surface
component in $a_{sym}(A)$ in addition to the mass-independent volume
component $a_{sym}^v$. In the literature, two different definitions
have been used for $a_{sym}(A)$. The first, hereafter referred to
as I \cite{dan} is,
\begin{eqnarray} 
a_{sym}(A)=\frac {a_{sym}^v}{1+\frac {a_{sym}^v}{\beta_E}A^{-1/3}}
\end{eqnarray} 
and the second, hereafter referred to as II \cite{sto} is,
\begin{eqnarray} 
a_{sym}(A)= a_{sym}^v-a_{sym}^sA^{-1/3}.
\end{eqnarray} 
In definition I, $\beta_E$ is a measure of the surface symmetry
energy, $a_{sym}^s$ is the surface symmetry energy coefficient
in definition II. In the limit of very large $A$,
$(a_{sym}^v)^2/\beta_E \sim a_{sym}^s$. The phenomenological
value of $a_{sym}^s$ is taken as $\sim $ 45 MeV \cite{rei,sto,jia}
and that of $a_{sym}^v/\beta_E$ is in the close range of $\sim $
2.4$\pm $0.4 \cite{dan,dan1,liu}.

It is evident that the symmetry energy coefficient has an extremely
important role in describing properly the nuclear binding energies
along the periodic table and in getting a broad understanding of
the nuclear drip lines. It also plays a seminal role in guiding
the dynamical evolution of the core collapse of a massive star
and the associated explosive nucleosynthesis. A large (small)
magnitude of $a_{sym}$ inhibits (accelerates) change of protons
to neutrons through electron capture \cite{ste,jan}. This change
in isospin asymmetry has its import in the nuclear equation of
state (EOS) and thus on the dynamics of the collapse and explosive
phase of a massive star.
 Matter in that phase is warm, it is therefore
essential to know with precision the thermal
dependence of the symmetry coefficients. Furthermore, in this collapse
or bounce phase, the nuclear matter is inhomogeneous; it 
nucleates to clusters of different sizes.
Knowledge about the thermal evolution of the symmetry
coefficients of finite nuclei then becomes  a matter of central importance.  

In the low temperature domain ($T \le $2 MeV), calculations of the symmetry 
coefficients of atomic nuclei have been done earlier by Donati {\it et.al}
\cite{dona} in a schematic model. The motion of the nucleons in a fluctuating
mean-field results in a nucleon effective mass that carry signatures of
nonlocality in space (the $k$-mass $m_k$) and also nonlocality in
time (the energy-mass $m_\omega $). The energy mass $m_\omega $ is seen
to decrease with temperature \cite{pra,has}, 
this brings in a decreased density of
states and thus an increase in the symmetry coefficient. Calculations
in this limited temperature range have further been done by Dean {\it et.al} 
\cite{dea} in a shell model Monte-Carlo framework. It provides qualitative
support to these earlier findings. The symmetry coefficients, however,
are found to be much below the nominally accepted values.
Evaluation of the temperature dependence of the volume and surface symmetry 
coefficients of nuclei have also recently been attempted by Lee and Mekjian
\cite{lee} in a density functional theoretic approach. These calculations
are also limited to low temperatures ($T \le $3 MeV); the approximations 
employed here keep the results meaningful in this small temperature domain. 

Exploring the thermal evolution of the symmetry coefficients of specific
atomic masses has  been attempted \cite{de2} 
in a broader temperature
range ($T \le $ 8 MeV) more recently. 
The energies and free energies of the hot nuclei
are calculated in the finite-temperature Thomas-Fermi framework (FTTF)
with the subtraction technique \cite{sur} with suitable choice
of effective interactions. Dynamical changes in the energy-mass $m_\omega $
are taken care of. For a nucleus of mass $A$, the symmetry coefficient
is defined as
\begin{eqnarray}
a_{sym}(A,T)= [e_n(A,X_1,T)-e_n(A,X_2,T)]/(X_1^2-X_2^2).
\end{eqnarray}
 Here $e_n$'s are the nuclear part of the energy per nucleon of the 
nuclear pair of mass $A$ but having different charges 
and $X_1$ and $X_2$ are the asymmetry parameters of the nuclei. 
For a finite nucleus with $Z$ protons and $N$ neutrons, $X$ is
defined as $(N-Z)/A$. Similar to $a_{sym}(A,T)$, 
the symmetry free energy coefficient $f_{sym}(A,T)$ can be defined.
These definitions suffer from the fact that unique values of $a_{sym}$ or
$f_{sym}$ for a nucleus of mass $A$ can not be prescribed; the values
depend on the choice of the isospin asymmetric nuclear pair.

The present communication is aimed to arrive at  unambiguous 
values of the temperature dependence of the symmetry 
coefficients. For a set (sixty nine)
of spherical and non-spherical nuclei covering almost the entire
periodic table (we take 36 $\le A \le 218$  and $ 14 \le Z \le 92$,
the list of the nuclei is taken from Ref. \cite{klu}), the energies
and free energies are calculated in the subtracted FTTF procedure,
taking into account the dressing of the nucleon mass to
energy-mass $m_\omega $ that arises from the coupling of the nucleonic
motion with the surface vibrations \cite{pra,has,shl}. Two effective
interactions are chosen, i) the zero-range Skyrme-type interaction
KDE0 \cite{agr} and ii) the finite-range modified Seyler-Blanchard
(SBM) interaction. The KDE0 interaction reproduces the binding
energies of many nuclei ranging from normal to exotic ones with a 
deviation which is much less than 0.5$\% $ for most cases. In addition,
it has been extremely successful in reproducing the breathing mode
energies of many nuclei, their charge radii and spin-orbit splitting.
The SBM interaction also has been very successfully applied in
getting properly the ground state binding energies \cite{mye1},
charge rms radii, giant monopole resonance energies etc. \cite{de3,maj}.
The SBM interaction is given by 

\begin{eqnarray}
v_{eff}(r, p,\rho )=C_{l,u} \Bigl [v_1(r,p)
+v_2(r,\rho) \Bigr ], \nonumber \\
v_1=-(1-p^2/b^2)f({\bf r}_1,{\bf r}_2), \nonumber \\
v_2=d^2 \Bigl [ \rho(r_1)+\rho (r_2) \Bigr ]^\kappa 
f({\bf r}_1,{\bf r}_2),
\end{eqnarray}
with
\begin{eqnarray}
f({\bf r}_1,{\bf r}_2)=\frac {e^{-|{\bf r}_1-{\bf r}_2|/a}}
{|{\bf r}_1-{\bf r}_2|/a}.
\end{eqnarray}

The strength parameters $C_l$ for like pairs (n-n,p-p) and $C_u$
for unlike pairs (n-p) carry information on the isospin dependence
in the interaction. The densities at the sites $\bf r_1$ and $\bf r_2$
of the two interacting nucleons with momenta $\bf p_1$ and $\bf p_2$
are given by $\rho (\bf r_1)$ and $\rho (\bf r_2)$; $\bf r =|\bf r_1
-\bf r_2 |$ and $\bf p =| \bf p_1 -\bf p_2 |$. The range of 
the interaction is $a$; $b$ and $d$ are measures of the momentum 
and density dependence in the interaction and $\kappa $ controls
the stiffness on the nuclear EOS. The procedures for determining 
these parameters are given in detail in Refs. \cite{ban,maj}. The
parameters for KDE0 and SBM interaction are listed in Table I and
II, respectively. The values of the saturation density $\rho_s$,
the volume energy,   
 the isoscalar volume incompressibility $K_\infty $, the
volume symmetry coefficient $a_{sym}^v$, the symmetry incompressibility
$K_{sym}$, the symmetry pressure $L$ and the critical temperature
\begin{table}
\caption{ The parameters of the KDE0 effective interaction }
\begin{ruledtabular}
\begin{tabular}{cccccccccc}
$t_0$& $t_1$& $t_2$& $ t_3$& $x_0$& $x_1$&$x_2$&$x_3$& $w_0$ & $\alpha$\\
     (MeV$\cdot$fm$^3$)& (MeV$\cdot$fm$^5$) &(MeV$\cdot$fm$^5$) &
 (MeV$\cdot$fm$^{3(1+\alpha)}$)&   & & & & MeV$\cdot$fm$^5$& \\
\hline
  -2526.52&430.94&-398.38&14235.52&0.7583&   -0.3087&-0.9495&
1.1445 &128.95& 0.1676 \\
\end{tabular}
\end{ruledtabular}
\end{table}
\begin{table}
\caption{ The parameters of the SBM effective interaction (in MeV fm units)}
\begin{ruledtabular}
\begin{tabular}{cccccc}
$C_l$&  $C_u$& $a$& $b$& $d$& $\kappa $\\
\hline
348.5& 829.7& 0.6251& 927.5& 0.879& 1/6\\
\end{tabular}
\end{ruledtabular}
\end{table}
$T_c$ for these two interactions are listed in Table III. 
It is worthwhile to note that the values of the symmetry 
coefficients $a_{sym}^v$, $K_{sym}$ and $L$ lie in the
range suggested by the empirical constraints emerging out of the analyses
of different recent experimental data \cite{she,roc,che,don}. The
method for obtaining the density profiles of hot nuclei and their
binding energies in the subtracted FTTF approach, with subsequent
modification due to energy-mass with the SBM and Skyrme-type interaction
has been described in some good detail in a recent article
\cite{de2}; we therefore do not repeat it here. 
The energies and free energies of the chosen sixty nine nuclei are
calculated with this prescription in a temperature grid. At a particular
temperature, the energies are then fitted in the framework of the
Bethe-Weiz\"acker mass formula
\begin{eqnarray}
E(N,Z,T)&=&a_v(T)A + a_s(T)A^{2/3} + a_c\frac {Z^2}{A^{1/3}} \nonumber \\
&&+a_{sym}(A,T)X^2A, 
\end{eqnarray}

\begin{eqnarray}
F(N,Z,T)&=&f_v(T)A + f_s(T)A^{2/3} + a_c\frac {Z^2}{A^{1/3}} \nonumber \\
&&+f_{sym}(A,T)X^2A.
\end{eqnarray}
In Eq.~(6), $a_v, a_s, a_c$ and $a_{sym}$ are the volume,
surface, Coulomb and symmetry energy coefficients. Similarly,
$f_v, f_s$ and $f_{sym}$ are the corresponding free energy coefficients.
The Coulomb energy has an implicit temperature dependence; it 
does not contribute to entropy. 
Since they are precisely known in a calculation, we try to
make a four-parameter fit with only the nuclear part of the energies and
free energies,
\begin{eqnarray}
E_n(N,Z,T)=a_v(T)A + a_s(T)A^{2/3} 
+a_{sym}(A,T)X^2A,
\end{eqnarray}

\begin{eqnarray}
F_n(N,Z,T)=f_v(T)A + f_s(T)A^{2/3} 
+f_{sym}(A,T)X^2A.
\end{eqnarray}
Here  $E_n$ and $F_n$ are the nuclear part of the
energy and free energy of the nucleus; $a_{sym}(A,T)$ is given by
Eq.~(1) or Eq.~(2). In a similar spirit, $f_{sym}(A,T)$ is written
as 
\begin{eqnarray}
f_{sym}(A,T)=\frac {f_{sym}^v(T)}{1+\frac {f_{sym}^v(T)}
{\beta_F (T)}A^{-1/3}}
\end{eqnarray}
or
\begin{eqnarray}
f_{sym}(A,T)= f_{sym}^v(T) -f_{sym}^s(T)A^{-1/3}.
\end{eqnarray}
The four-parameter set $f_v,f_s,f_{sym}^v$ and $f_{sym}^s$ (or 
$\beta_F$) have the same connotation as the set $a_v, a_s, a_{sym}^v$
and $a_{sym}^s$ (or $\beta_E$), except that the former set refers to
free energy.
The parametric values of the volume energy $a_v$ and the volume symmetry
free energy $f_v$ are shown as a function of temperature in panels (a) and
(b) of Fig.~1. At $T$=0, $a_v$ (or $f_v$) very closely reproduces the
energy per nucleon of symmetric nuclear matter.
\begin{figure}
\includegraphics[width=1.0\columnwidth,angle=0,clip=true]{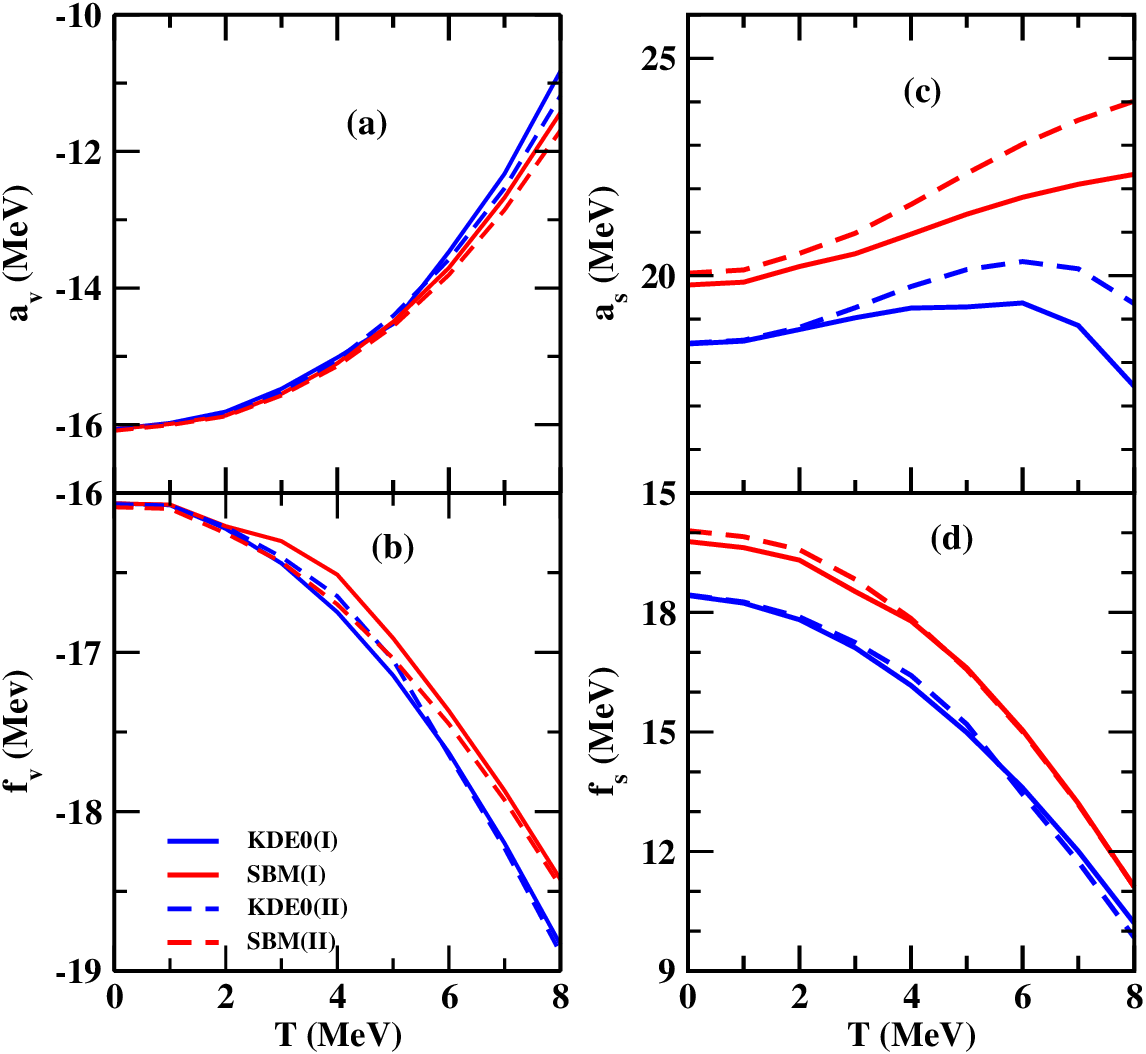}
\caption{(color online) The volume energy and volume free energy
coefficients shown as a function of temperature
in the left panels (a) and (b).
The full and dashed lines correspond to parametrization I and
II, respectively. The blue color refers to the KDE0 interaction,
the red color refers to SBM. In the right panels (c) and (d),  
the thermal dependence of surface energy and surface free energy
is shown. }
\end{figure}
At low temperatures, $a_v$ and $f_v$ are nearly
independent of the interactions chosen, at higher temperatures, a slight
dependence is observed. For a particular interaction, these values, however,
do not show any significant dependence on the chosen  set I or II. 
Both $a_v$ and $f_v$ are seen to change quadratically with temperature.
They are very well approximated with $a_v(T) = e(T=0) + T^2/K_1$
and $f_v(T) = f_v(T=0) - T^2/K_2 $, with $K_1 \sim $ 15.5 MeV and
$K_2 \sim $ 24.0 MeV. It is to be noted that for infinite matter at a
particular density and temperature, the energy and free energy
are canonically related (the entropy $S=-(\partial F/\partial T)_{\rho }$,
whence $K_1=K_2$); in the present case, density is a varying profile, also 
$a_v(T)$ and $f_v(T)$ are obtained from a least-squares fit
to the energies of a multitude of nuclei. This may explain the
different values of $K_1$ and $K_2$.

\begin{table}
\caption{ Some  bulk properties for infinite nuclear matter at the
saturation density for  the KDE0 and SBM effective interactions.
The values of saturation density are in fm$^{-3}$ and all other quantities
are in MeV.
 }
\begin{ruledtabular}
\begin{tabular}{cccccccc}
Force &$\rho_s$& $a_v$& $K_\infty$ & $K_{sym}$& $a_{sym}^v$ & $L$& $T_c$ \\
\hline
 KDE0 &0.161&-16.1 &229 &-144&  33 & 45.2 & 14.7\\ 
 SBM &0.154 &-16.1   & 238 & -101&31 & 59.8 & 14.9\\
\end{tabular}
\end{ruledtabular}
\end{table}
In the right panels (c) and (d) of Fig.~1, 
the thermal evolution of surface energy and the surface 
free energy coefficients are shown. 
The surface energy (upper panel)
increases slowly with temperature; with the KDE0 interaction, a 
slight fall at very high temperatures is, however, observed. With
temperature, the surface free energy (lower panel) decreases. In the 
literature \cite{lev,bon}, different  parametric forms  for the 
dependence of surface free energy on temperature have been used. We 
find that the form  of the type
$f_s=f_s(T=0) [(T_c^2-T^2)/(T_c^2+T^2)]^\alpha $ used in Ref. \cite{bon}
gives a reasonably good fit with our calculated values for both the
interactions using both the parameter set I and II
with $\alpha \sim $~0.95.

\begin{figure}
\includegraphics[width=1.0\columnwidth,angle=0,clip=true]{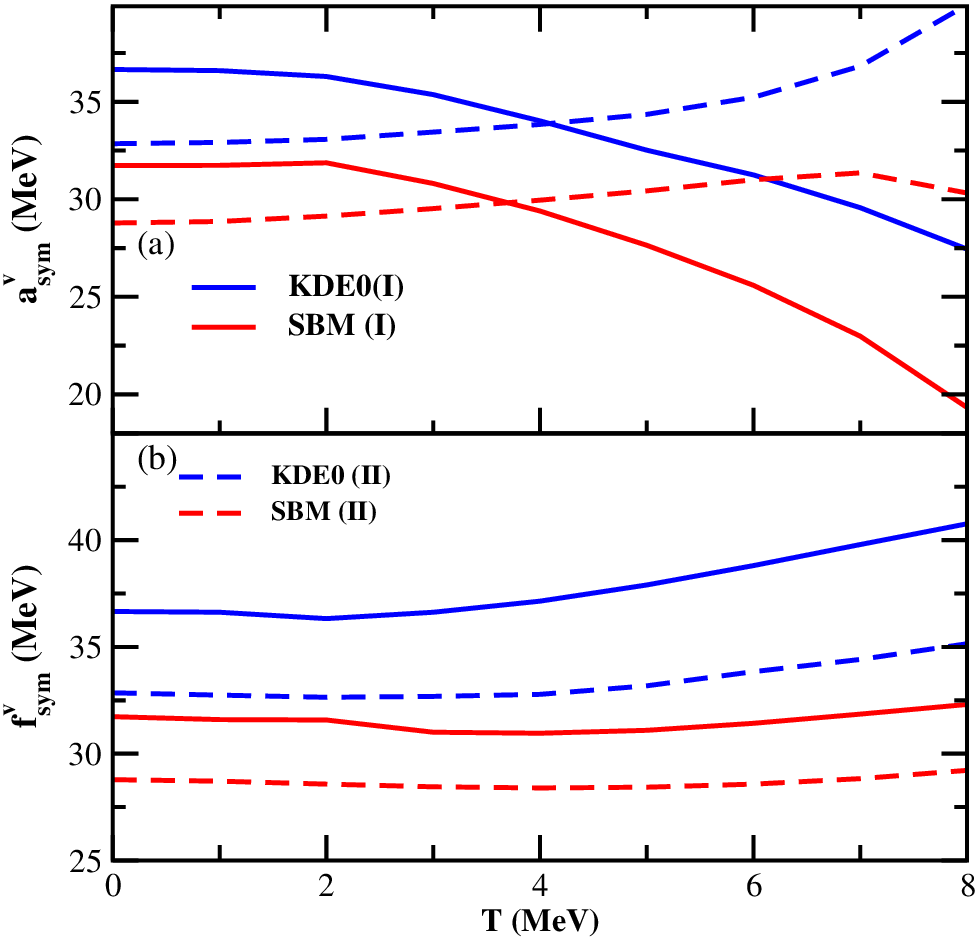}
\caption{(color online) The volume symmetry energy and the volume
symmetry free energy coefficients plotted as a function of temperature.
The lines and the colors have the same meaning as in Fig.~1.   }
\end{figure}

In Fig.~2, the evolution of the volume symmetry energy $a_{sym}^v$ and
the volume symmetry free energy $f_{sym}^v$ coefficients with temperature
are displayed in panels (a) and (b), respectively. The behavior of
$a_{sym}^v$ depends on how $a_{sym}(A)$ is defined. In definition I,
it falls with temperature, in definition II, it shows a slow increase.
The nature of the fall of $a_{sym}^v$ (in I) or its increase (in II)
is nearly the same for both the interactions. The coefficient $f_{sym}^v$,
however, shows nearly no dependence on 
temperature for both the interactions and 
in both definitions.

In the left panels of  Fig.~3, the thermal 
dependence of the coefficients $\beta_E$ and $\beta_F$
as used in Eqs~(1) and (10) in the definition I of $a_{sym}(A)$ and
$f_{sym}(A)$ is shown. 
\begin{figure}
\includegraphics[width=1.0\columnwidth,angle=0,clip=true]{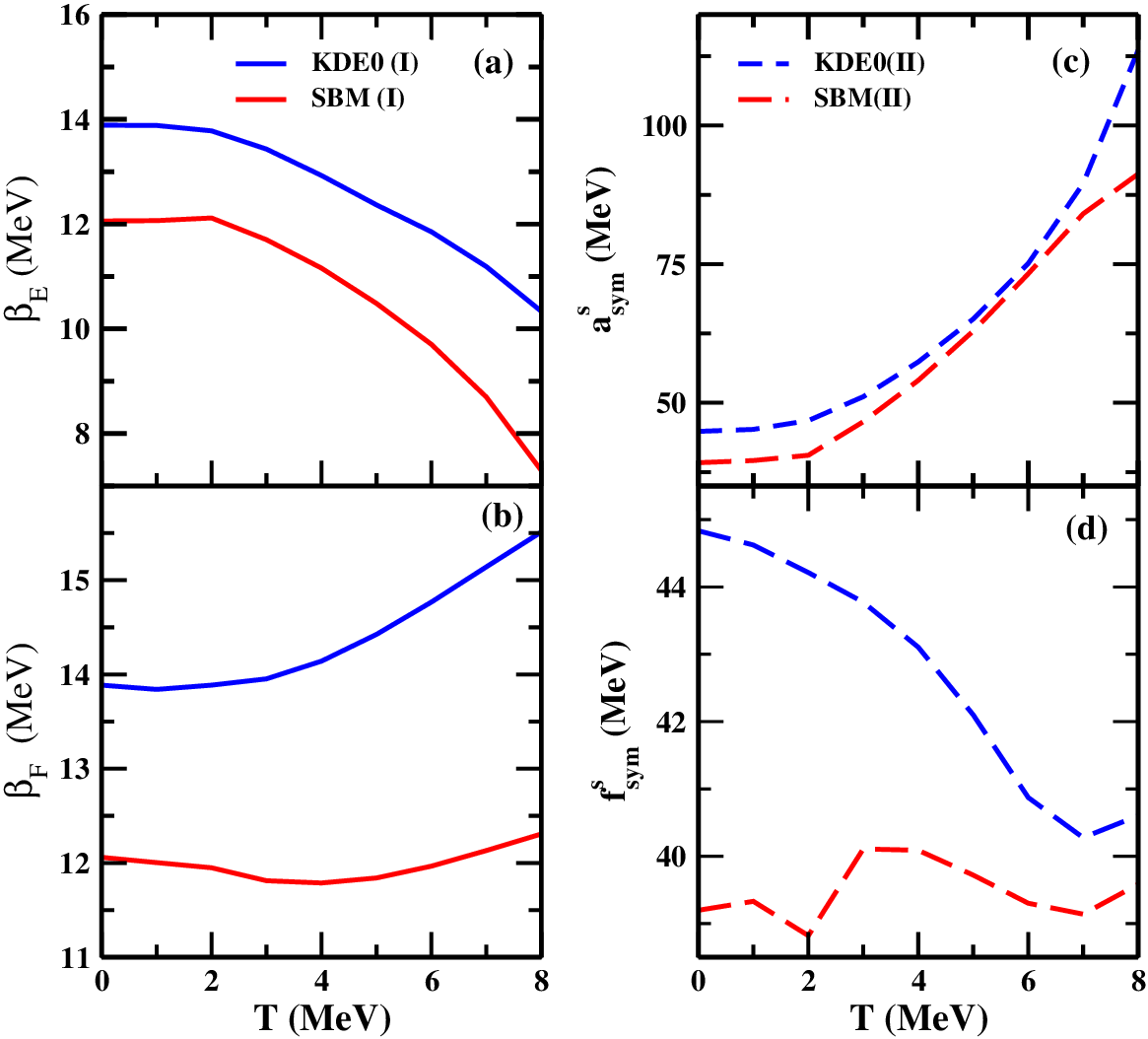}
\caption{(color online) The thermal dependence of the coefficients
$\beta_E$ and $\beta_F$ (from parametrization I) is displayed
in panels (a) and (b).
The thermal evolution of the surface symmetry
coefficients $a_{sym}^s$ and $f_{sym}^s$ (from parametrization II)
is shown in the right panels (c) and (d). 
The blue and red lines in both the left and right panels
 refer to calculations
with the KDE0 and SBM interactions, respectively.  }
\end{figure}
At $T=0$, the value of $\beta_E$ or $\beta_F$ is
12.1 and 13.9 MeV for the SBM and KDE0 interactions, respectively;
they compare well with the value  of $\sim $ 13 MeV 
obtained from analyses of the 'experimental'
symmetry energies of isobaric nuclei \cite{liu}. With temperature, $\beta_E$
decreases for both the interactions; $\beta_F$ shows a nominal increase.
We have, however, noticed that both $a_{sym}^v/\beta_E$ and
$f_{sym}^v/\beta_F$ are nearly temperature independent, lying
in the range of $\sim $ 2.64$\pm $0.01.

The temperature-dependent surface symmetry coefficients $a_{sym}^s$
and $f_{sym}^s$ as used in Eqs.~(2) and (11) in the definition II
are shown in  the right panels of Fig.~3. 
\begin{figure}
\includegraphics[width=1.0\columnwidth,angle=0,clip=true]{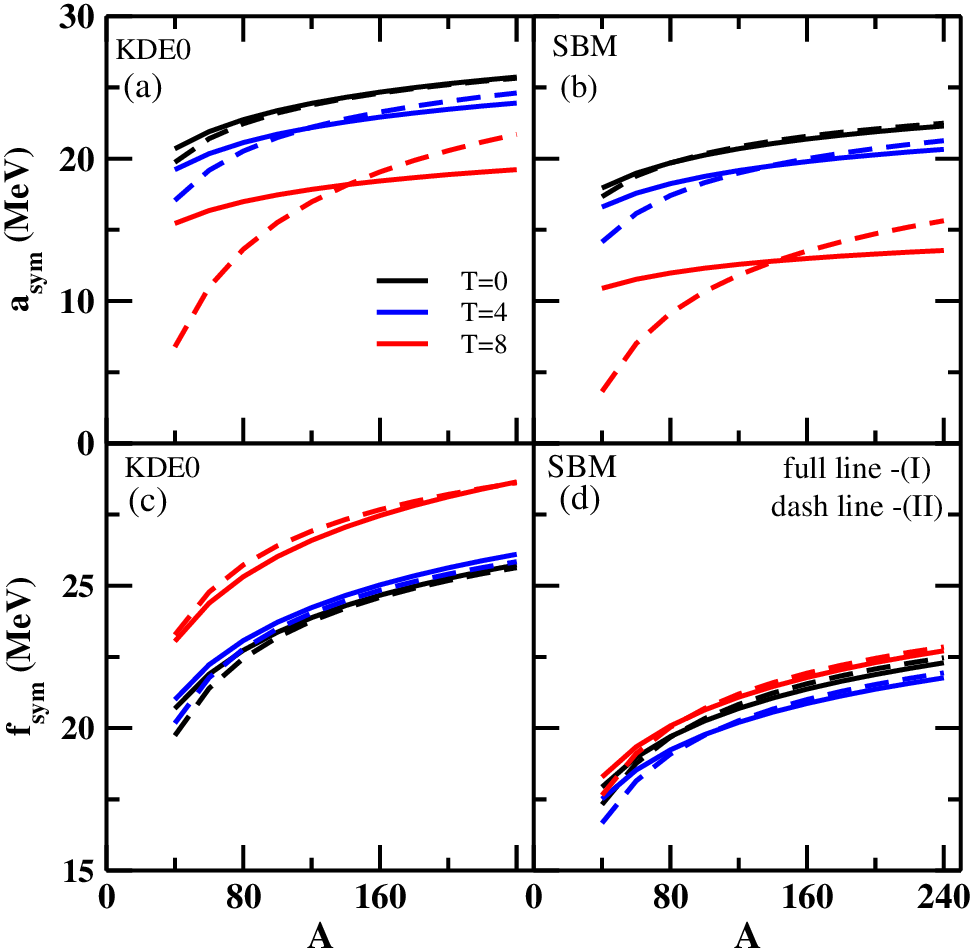}
\caption{(color online) The symmetry coefficients $a_{sym}(A)$
and $f_{sym}(A)$ shown as a function of mass number for three
temperatures. The full and dashed lines refer to parametrization I
and II. The black, blue and red lines correspond to $T$ =0, 4, and 
8 MeV, respectively.  }
\end{figure}
At $T=0$, $a_{sym}^s$ is 44.8 MeV and 39.2 MeV
for the KDE0 and SBM interactions, respectively, close to the
phenomenological value of $\sim $ 45 MeV \cite{sto}. With temperature, for 
both the interactions, $a_{sym}^s$ increases sharply showing the
growing importance of the surface term in $a_{sym}(A)$. The surface  free
energy coefficient $f_{sym}^s$, however, displays a slow decrease with
temperature for the KDE0 interaction. As for the SBM interaction, 
$f_{sym}^s$ is nearly temperature-independent.

A comparison with calculations in Ref. \cite{lee} may now be in order.
In both calculations, the surface symmetry coefficient seems to be 
more sensitive to temperature compared to the volume symmetry
coefficient. However, in Ref. \cite{lee}, in the limited temperature
range they explore, the temperature dependence of the
surface coefficients seem to be more pronounced than those seen in
the present calculation. There are subtle differences too, the
lack of self-consistency of the density profiles used in \cite{lee}
alongwith the low-temperature, high-density approximations involved
may be the reason behind these differences.

In Fig.~4, the mass dependence of the $a_{sym}(A)$ and $f_{sym}(A)$
is shown at three temperatures, $T=$ 0, 4 and 8 MeV. 
\begin{figure}
\includegraphics[width=1.0\columnwidth,angle=0,clip=true]{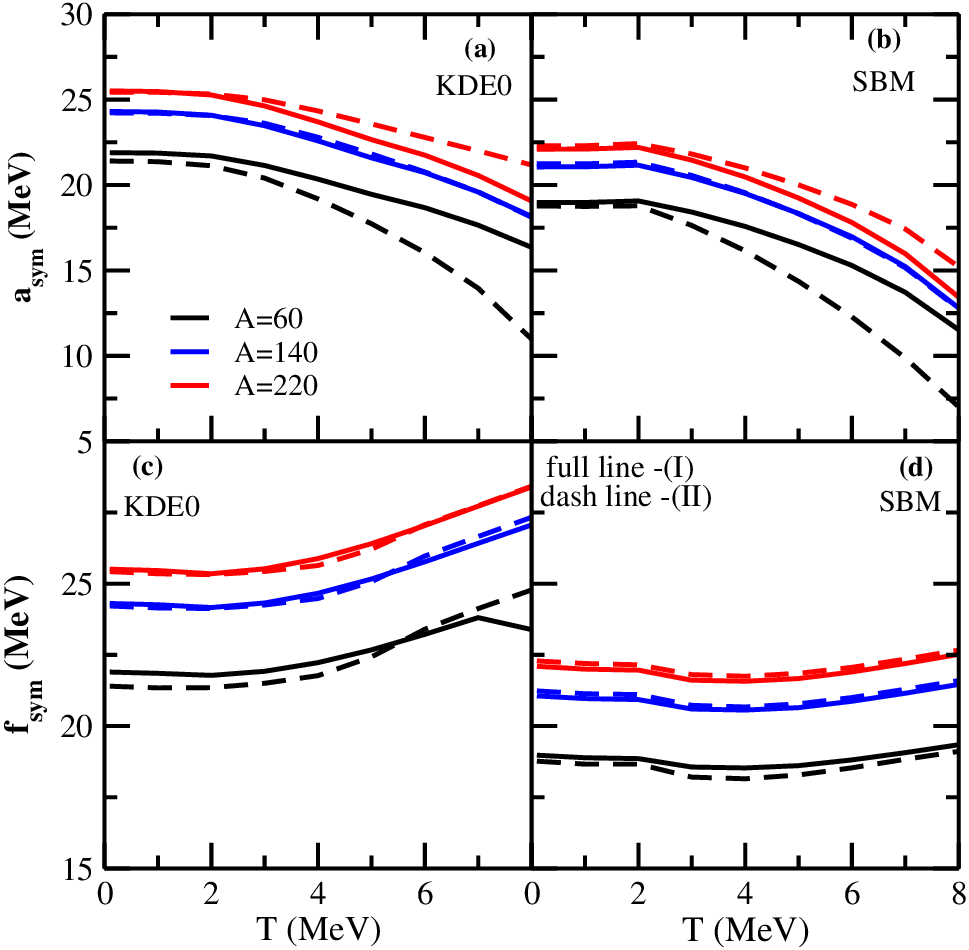}
\caption{(color online) The thermal dependence of the symmetry coefficients
of nuclei shown for three mass number. The full and dashed lines
refer to parametrization I and II. The black, blue and red
lines correspond to $A$ =60, 140 and 220, respectively. 
}
\end{figure}
Panels (a) and (b) in the figure display $a_{sym}(A)$ for KDE0 and SBM
interactions, respectively; panels (c) and (d) display $f_{sym}(A)$.
The full lines correspond to definition I for the symmetry coefficients,
the dashed lines do the same for definition II. The general findings
are : for a particular mass number, $a_{sym}(A)$ decreases with
temperature, $f_{sym}(A)$ increases. At fixed temperature, $a_{sym}(A)$
and $f_{sym}(A)$ increase with $A$; this follows from the definitions. The
values of $f_{sym}(A)$ seem to depend little on the parametrization
I or II;
similar is the case  for $a_{sym}(A)$ except at very high temperature.

For the fixed values of nuclear masses, the temperature dependence
of $a_{sym}(A)$ and $f_{sym}(A)$ are exhibited in Fig.~5. The masses
chosen are $A=$ 60, 140 and 220. Panels (a) and (b) in this figure
display $a_{sym}(A)$ for KDE0 and SBM interactions, respectively;
panels (c) and (d) do the same for $f_{sym}(A)$. The black lines
pertain to $A=$ 60, the blue lines to $A=$ 140  and the red lines
correspond to $A=$ 220. The general findings in Fig.~4 that $a_{sym}(A)$
falls and $f_{sym}(A)$ shows a very slow 
increase with temperature is reinforced
from this figure. 
For the SBM interaction, a near constancy of $f_{sym}(A)$
with a slight dip in the middle of the temperature range is seen. This
was also occasionally observed earlier \cite{de3} with a different 
definition of $f_{sym}(A)$ - in the spirit of Eq.~(3). It is also
observed that for a chosen interaction, both parametrization I and II
yield nearly the same value of the symmetry coefficients 
except for $A=$ 60 at higher temperatures.  

To summarize, in a liquid-drop-model-inspired fit of the total energies
and free energies of a system of nuclei evaluated in a subtraction-implemented
finite temperature Thomas-Fermi framework, the temperature dependence
of the symmetry energy coefficients of nuclei have been 
evaluated in this communication. 
Two different energy density functionals, one with the zero-range
Skyrme-type KDE0 and the other with a finite-range SBM interaction
have been employed for this purpose. The general behavior of the
temperature dependence of the symmetry coefficients seems to be
nearly independent of the energy functional used.
For cold systems, the calculated volume and surface 
symmetry energy coefficients 
lie within the constraints set
from  analyses of different experimental data. With temperature,
the symmetry free energy coefficients show a weak change.
A strong temperature dependence of $a_{sym}^v$ is however observed,
the temperature dependence of $a_{sym}^s$ is even stronger; this
results in a rapid fall in $a_{sym}(A)$ of the atomic nucleus 
as the temperature rises. 
The calculations, in addition  throw light on the 
thermal mapping of the volume
and surface energies which are in excellent qualitative agreement
with those in common usage.

\begin{acknowledgments}
 J.N.D  acknowledges support of DST, Government of India.
\end{acknowledgments}

\end{document}